# An improved distributed routing algorithm for Benes based optical NoC

Jing Zhang*[a], Huaxi Gu[a], Yintang Yang[b]
[a]State Key Laboratory of Integrated Service Networks, Xidian University, Xi'an, China,
[b]Institute of Microelectronics, Xidian University, Xi'an, China

## ABSTRACT

Integrated optical interconnect is believed to be one of the main technologies to replace electrical wires. Optical Network-on-Chip (ONoC) has attracted more attentions nowadays. Benes topology is a good choice for ONoC for its rearrangeable non-blocking character, multistage feature and easy scalability. Routing algorithm plays an important role in determining the performance of ONoC. But traditional routing algorithms for Benes network are not suitable for ONoC communication, we developed a new distributed routing algorithm for Benes ONoC in this paper. Our algorithm selected the routing path dynamically according to network condition and enables more path choices for the message traveling in the network. We used OPNET to evaluate the performance of our routing algorithm and also compared it with a well-known bit-controlled routing algorithm. ETE delay and throughput were showed under different packet length and network sizes. Simulation results show that our routing algorithm can provide better performance for ONoC.

## 1. INTRODUCTION

Previous systems were mostly based on SOC, which uses shared-bus to process data. With the tremendous desire on more computational complexity when using different new technologies, such as the limited feedback technique in HSDPA and LTE systems [1] and the recent research on interference alignment for cellular systems [2, 3], network on chip (NOC) has drawn much of the attentions. To meet the requirements of sub-micron technology, the new design concept, so called Network-on-Chip (NoC), has been proposed as a long-term and ideal solution to solve some serious bottleneck problems that traditional SoC had faced. NoC is an alternative approach that uses packet switching to communicate between on-chip components. NoC endeavors to bring network communication methodologies to on-chip communication to address problems like performance, reliability, scalability and power consumption for architecture design [4]. But it's becoming increasingly hard for electrical NoC to satisfy the design requirements of power, delay and bandwidth. Integrated optical interconnect is believed to be one of the main replacement technologies. So more and more researches of using optical interconnects to replace electrical wires come into being, hence a a novel concept-ONoC (optical NoC). ONoC is accepted to enable ultra-high throughput, low power dissipation, minimal access latencies and high bandwidth [5].With recent advances in optical devices, this design methodology can provide scalable and reliable data transmission and reduce the area and power consumption on processors for on-chip communication. The choice of topology, switching mechanism, flow control mechanism and routing algorithm are still the major challenges that ONoC designers are facing nowadays.

Several topologies had been proposed and discussed extensively for ONoC communication. In this letter, we choose the well-known Benes network as the topology for ONoC application. The Benes network has several superior features, such as easy scalability, symmetry and low degree. What's more, multistage interconnection pattern enables Benes ONoC communication to have the capability to avoid deadlock and live lock, which is an obviously advantage unlike other mesh-based topologies. When the network size is enlarged, Benes ONoC architecture still could maintain good performance.

An important problem of ONoC design is designing the routing algorithms. Several efficient routing algorithms have been proposed for Benes network in previous researches [6-8]. But all these routing algorithms are designed for permutation assignments, which are not applicable for Benes ONoC communication. ONoC need a distributed algorithm to determine a reasonable path for any given pair of nodes. As it known to all, the distributed routing algorithms can be classified into non-adaptive and adaptive categories according to the degree of adaptiveness provided by algorithms. In this letter, we propose a new distributed routing algorithm with adaptiveness for Benes ONoC. Furthermore, we compare the performance of our routing algorithm with the famous bit-controlled algorithm [8]. Simulation results demonstrate that the new distributed routing algorithm enables Benes network to realize on-chip communication with lower latency and higher throughput.

The rest of paper is organized as follows. Section II describes the Benes-based ONoC architecture in details. We also propose a new distributed routing algorithm for Benes-based ONoC. Section III evaluates and analyzes the performance of our designed routing algorithm using a simulator called Opnet. The End to End delay and throughput are compared with a known bit-controlled routing algorithm under different packet length and network sizes. Conclusions are given in the last section.

## 2. ROUTING IN BENES ONOC ARCHITECTURE

### 2.1 Benes ONoC Architecture

Benes network is a well-known multistage network, which has received much attention during the last four decades [9-13]. An NxN Benes network, where N=$2^k$, consists of $2\log_2 N-1=2k-1$ stages, and each stage contains N/2 switching elements. As is shown in Fig. 1, $R_{ij}$ indicates that this router is the $i_{th}$ ($\forall i\in\{0,N/2-1\}$) switching element at the $j_{th}$ ($\forall j\in\{1,2k-1\}$) stage. The degree of this topology is 2. So the number of hops is the same for any routing path in Benes network.

From a functional point of view, in an NxN Benes-based ONoC, each source port $S_i$ ($\forall i\in\{0,N-1\}$) consists of a network interface (NI) and transmitter, and each destination port $D_j$ ($\forall j\in\{0,N-1\}$) consists of a receiver and NI. Message is sent from each source port through some routers to one or more destination ports depending on the routing algorithm. And the number N represents that the number of IP cores in a Benes-based ONoC architecture is N. Routers are responsible for the transmission of message stream from source to destination. What's more, each router in Benes-based ONoC architecture is a $2\times2$ basic switching element in nature. And $2\times2$ switching element has two states: cross state and bar state [14].Figure 1 shows an example of a $16\times16$ Benes-based ONoC architecture.

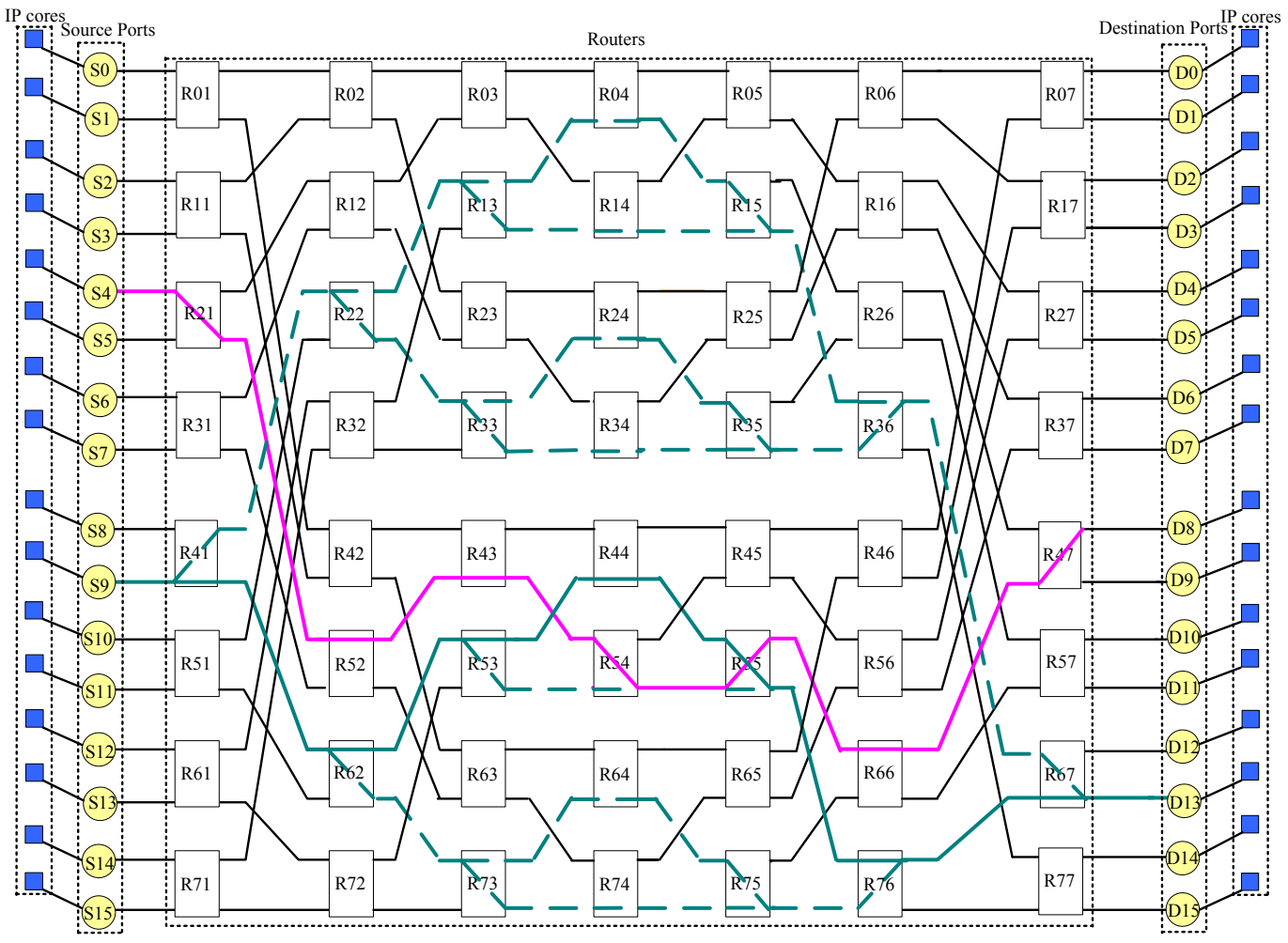

Figure 1. $16\times16$ Benes-based ONoC architecture and examples of messages routed in this network using partially routing algorithm.

### 2.2 Routing in Benes ONoC

Routing algorithm plays an important role in the efficiency of the ONoC communication which is used to determine an appropriate path that messages travel from a given source node to a given destination node. Adaptiveness is an important aspect for routing algorithm. Adaptive routing can help to bypass the congested or faulty nodes, hence improving performance of the network. In non-adaptive routing, which is also called deterministic routing, there is only one unique and predetermined path between any pair of source and destination node. The path is determined by the source and destination address without considering network state.

Several routing algorithms have been developed for Benes network to realize permutation assignments. A well-known bit-controlled routing algorithm (BCRA) [8] provides a unique predetermined path between any given pair of source and destination nodes. Routing of packets in BCRA depends on the routing label, which could be obtained from the destination address, contains $2k-1$ bits. The value of routing label is used to select the output port at eace stage without considering the network state. Taking into account the congestion in Benes-based ONoC, we propose a new distributed routing algorithm (DRA) which is also a adaptive algorithm. Our routing algorithm provides various routing choices for messages that traveling in the network. It takes into account the network state when making routing decisions, thus increasing the success rate of connection. Messages could select a proper path to be routed to the final destination and reduce the possibility of congestion at the same time.

Before describing our routing algorithm, assumption and definition are given below:

***Assumption:*** Messages are routed from source port $S_i$ to destination port $D_j$ ($j \neq i$). Denote the lower output port of router as $p_{lo,}$ and the upper output port as $p_{uo}$. Then $p_{so}$ represents the selected output port of router when making routing decisions.

***Definition:*** Routing of messages in stage 2k-b ($1 \le b \le k$) depends on the routing bit ($R_{bit}$), which can get from destination terminal $D_i$ by Formula (1).

$$R_{\text{bit}} = \left\lfloor i / 2^{b-1} \right\rfloor_{MOD2} \quad (1)$$

Operation of the routers in stage 2k-b depends on the value of the routing bit. If the value is zero, message is routed to the upper output port of the switching node, otherwise to the lower output port.

The distributed routing algorithm is described as follows.

```
/*The distributed algorithm for N×N Benes-based ONoC */
/*Current router: R_ij*/
/*Source node: S_m*/
/*Destination node: D_n*/
/* N = 2^k */
/*b (1 ≤ b ≤ k ) is an integer*/
Begin
if (j>=1&&j<k&& p_uo is busy)
            p_so = p_lo of R_ij;
  else if (j>=1&&j<k&& p_lo is busy)
            p_so = p_uo of R_ij;
  else if (j>=1&&j<k&& p_lo and p_uo are free)
            p_so = p_lo or p_uo of R_ij;
  else if (j==(2k-b))
       {
         if (R_bit==0)
            p_so = p_uo of R_ij;
          else if (R_bit==1)
            p_so = p_lo of R_ij;
       }
Return output port p_so
End
```

***Examples:*** We assume that a connection from $S_4$ to $D_8$ already exists in Benes-based ONoC architecture. For a new message, which arrives at $S_9$ and goes for $D_{13,}$ there are 8 possible paths can be selected as shown in dashed line according to the proposed distributed routing algorithm. If deterministic routing algorithm is used, such message may be blocked at $R_{54,}$ because the lower output port of $R_{54}$ is occupied by connection from $S_4$ to $D_8$. However, our routing algorithm can choose the other path in $R_{53}$ for the message. Hence, a new connection may be established as the blue solid line shows.

## 3. PERFORMANCE EVALUATION

### 3.1 Simulation Environment

In order to evaluate the new routing algorithm in the Benes ONoC, we apply the simulator OPNET to construct Benes ONoC models and use circuit switching technique as the communication mechanism. When a connection is needed to be established, a path-setup packet carrying the source address and the destination address and some other control information is first sent to select routing path. At the same time, a sequence of routing links is reserved. When the path-setup packet reaches the destination, a path has been constructed and an acknowledgement (ACK) packet is sent back to the source node along the constructed path. If the source node has received ACK, packet could begin to be sent along the reserved path. In simulation, the bandwidth of each channel is assumed to be 12.5 Gbps which can be achieved by nanophotonic device nowadays [5]. Messages are generated following the passion process and the path-setup message is set to be 32 bytes. The uniform traffic pattern is employed which means the choice of destination address follows uniform distribution.

The performance of routing algorithm is measured by the aspects of End to End (ETE) delay and throughput. ETE delay is defined to be the average of the total delay of all messages generated in the network. Throughput is used to measure the rate of messages which are received by each destination node under a given offered load. We define ETE delay, throughput and offered load $\lambda$ as following equations (2), (3), (4) respectively.

$$Delay = \sum_{i=1}^{N_{received}} t_i / N_{received} \tag{2}$$

and

$$Throughput = \lambda \times N_{received} / N_{sent} . \tag{3}$$

and

$$\lambda = T_{transmission} / \left( T_{transmission} + T_{interval} \right) . \tag{4}$$

Where $t_i$ is the delay of the $i_{th}$ message, $N_{received}$ and $N_{sent}$ are the number of total messages that received by each destination node and sent by the source node respectively. $T_{transmission}$ represents the time used to transmit messages while $T_{interval}$ is the arrival time gap between two messages. $T_{interval}$ follows exponentially distribution.

### 3.2 Simulation Results

We compare our distributed routing algorithm (DRA) with the bit-controlled routing algorithm (BCRA) through simulation. ETE delay and throughput are comparatively analyzed in $16\times16$ Benes network with the message length is set to be 32 bytes, 64 bytes and 128 bytes. As the offered load increases, the network enters the saturation state. But the value of saturation point is larger while using DRA (see Fig. 2(a)). Also we can find, value of saturation point is more and more large and ETE delay is lower at the same given offered load as the message length increases. The reason is that, DRA has more choices when network condition changes. And the proportion of the time used to transmit larger message increases. All these can help to reduce the blocking probability of the path-setup message. As a result, more messages could be received successfully by the destination. Hence, as is shown in Fig. 2(b), the throughput increases.

Simulations are also implemented on $8\times8$, $16\times16$ and $32\times32$ Benes network with message length of 64 bytes, which aims to compare the performance of DRA and BCRA when network size enlarges. The results are shown in Fig. 3. It's easy to find the performance of DRA is better than BCRA when the network size enlarges. The larger network size can lead to more time consumption and larger blocking possibility. More messages will be generated and the contention will be more intense. As a result, the ETE delay increases and the throughput decreases. DRA has more choices than BCRA when contention occurs and can bypass the occupied router, hence reducing the blocking probability of messages. The improvement of DRA is obvious on latency and throughput of Benes network with different network size. For example, in $32 \times 32$ Benes network, average latency is decreased by 22.6183% and average throughput is improved by 21.6087% over BCRA for message size of 32 bytes. Compared with a partially adaptive routing algorithm proposed in [15], this DAR has been proved to improve slightly in terms of delay and throughput.

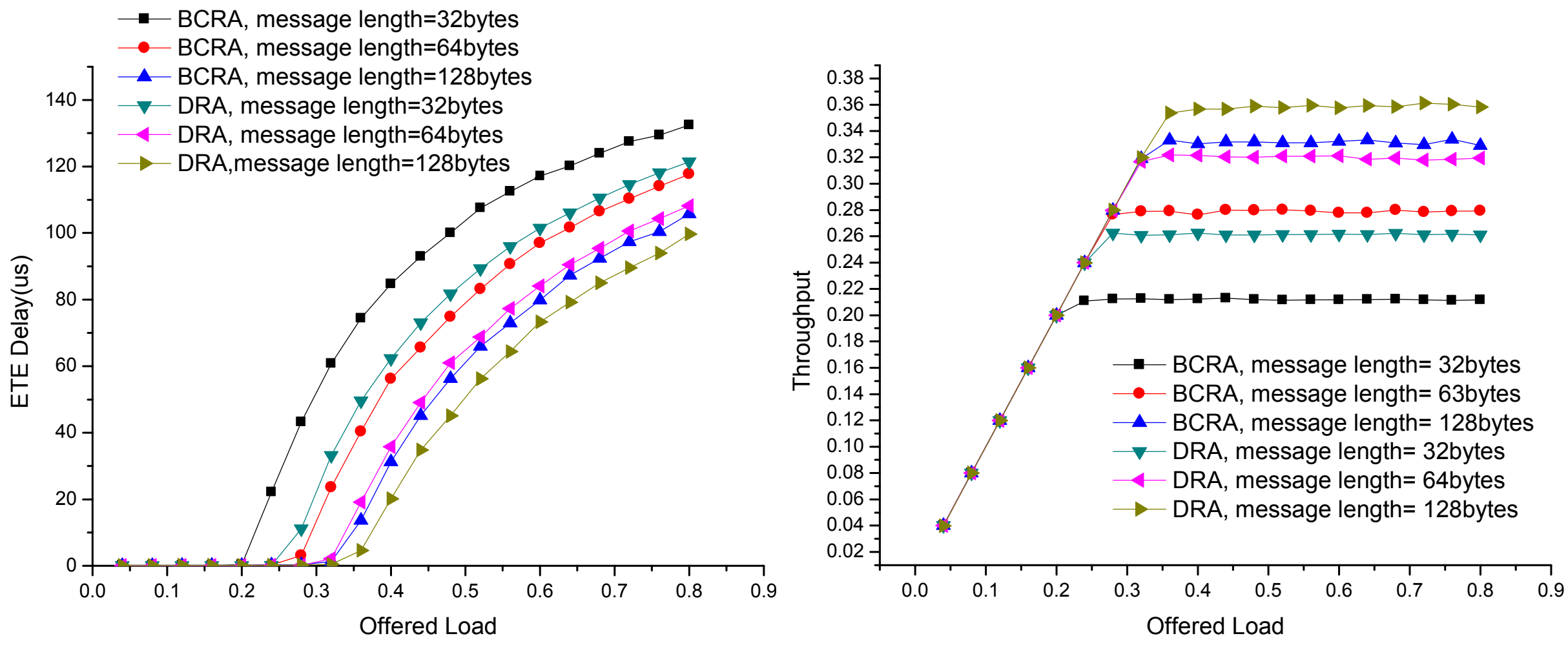

(a) ETE Delay (b) Throughput

Figure 2. Performance of a $16 \times 16$ Benes ONoC with different message lengths.

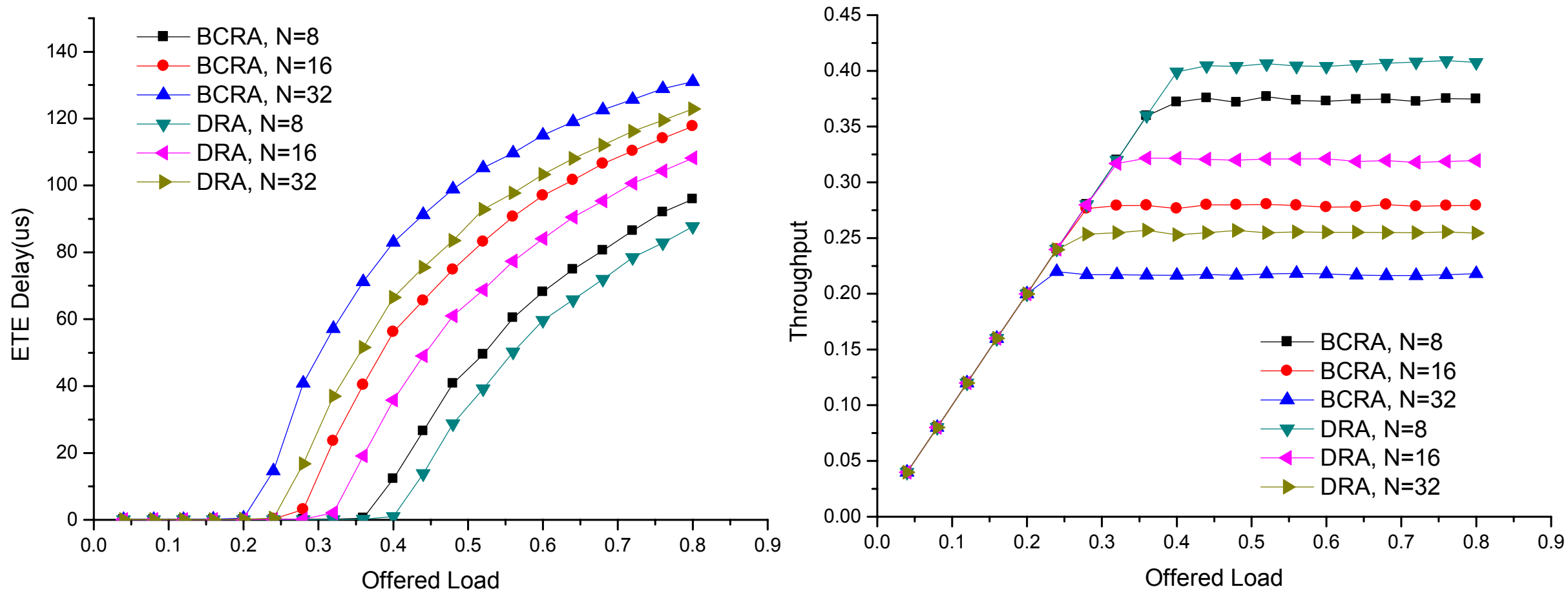

(a) ETE Delay (b) Throughput

Figure 3. Performance of Benes ONoC with different network sizes and messages length of 64 bytes.

## 4. CONCLUSION

In this letter, we propose an efficient distributed routing algorithm for Benes ONoC communication. Unlike traditional routing algorithms developed for Benes network, the new distributed routing algorithm can provide more choices for routing path between a given pair of source and destination node. It takes into account the network state and enables the routing messages to bypass the congested or faulty nodes. Then we apply simulator OPNET to evaluate the new routing algorithm from the aspects of ETE delay and throughput. The results demonstrate that the distributed routing algorithm can obtain lower ETE delay and higher throughput and could provide on-chip communication with higher reliability and efficiency.

## ACKNOWLEDGEMENT

This work is supported in part by the National Science Foundation of China under Grant No. 60803038, 60725415 and 60971066, the special fund from State Key Lab (No.ISN090306), and the 111 Project under Grant No. B08038.